\documentclass[12pt,fullpage]{article}

\usepackage[utf8]{inputenc}
\usepackage[T1]{fontenc}
\usepackage{amsfonts}
\usepackage{amsmath}
\numberwithin{equation}{section}
\usepackage{amssymb}
\usepackage{graphicx,subfigure}
\usepackage[T1]{fontenc}
\usepackage{graphics}
\usepackage{dsfont}
\usepackage{cases}
\usepackage{float}
\usepackage[title,titletoc,toc]{appendix}

\newcommand{\ba}{\begin{eqnarray}}
\newcommand{\ea}{\end{eqnarray}}

\newtheorem{theorem}{Theorem}[section]

\newtheorem{definition}[theorem]{Definition}

\newtheorem{lemma}[theorem]{Lemma}

\newtheorem{proposition}[theorem]{Proposition}
\newtheorem{remark}[theorem]{Remark}
\newenvironment{Proofc}[1]{\smallskip\par\noindent\textsc{#1}\quad}%
  {\hfill$\Box$\bigskip\par}

\title{Explicit phase diagram\\ 
for a one-dimensional blister model}
\author{
\normalsize\textsc{ G. Chmaycem$^{b,*}$, M. Jazar
\footnote{LaMA-Liban, Lebanese University, P.O. Box 37 Tripoli,
Lebanon. E-mail: ghada.chmaycem@gmail.com (G. Chmaycem),
mjazar@laser-lb.org (M. Jazar).
\newline \indent $\,\,{}^{b}$ Universit\'{e} Paris-Est,
CERMICS, Ecole des Ponts ParisTech, 6 et 8 avenue Blaise Pascal,
Cit\'e Descartes Champs-sur-Marne, 77455 Marne-la-Vall\'ee Cedex 2,
France. E-mail: chemaycg@cermics.enpc.fr (G. Chmaycem),
monneau@cermics.enpc.fr (R. Monneau)}  and R. Monneau$^{b}$}}

\begin{document}

\maketitle

\begin{center}
{\small{\textbf{Abstract} }}
\end{center}
We consider a thin film bonded to a substrate. The film acquires a residual stress upon cooling because of the mismatch of thermal expansion coefficient between the film and the substrate. The film tends to lift off the substrate when this residual stress is compressive and large enough. In this work, this phenomenon is described by a simplified one-dimensional variational model. We minimize an energy and study its global minimizers. Our problem depends on three parameters: the length of the film, its elasticity and a thermal parameter. 
Our main result consists in describing a phase diagram depending on those parameters in order to identify three types of global minimizers: a blister, a fully delaminated blister and a trivial solution (without any delamination).  Moreover, we prove various qualitative results on the shape of the  blisters and identify the smallest blister which may appear. 
\\




\section{Introduction}

\subsection{Physical motivation}

\begin{figure}[htb]
\begin{center}
\includegraphics[width=1.02\textwidth]{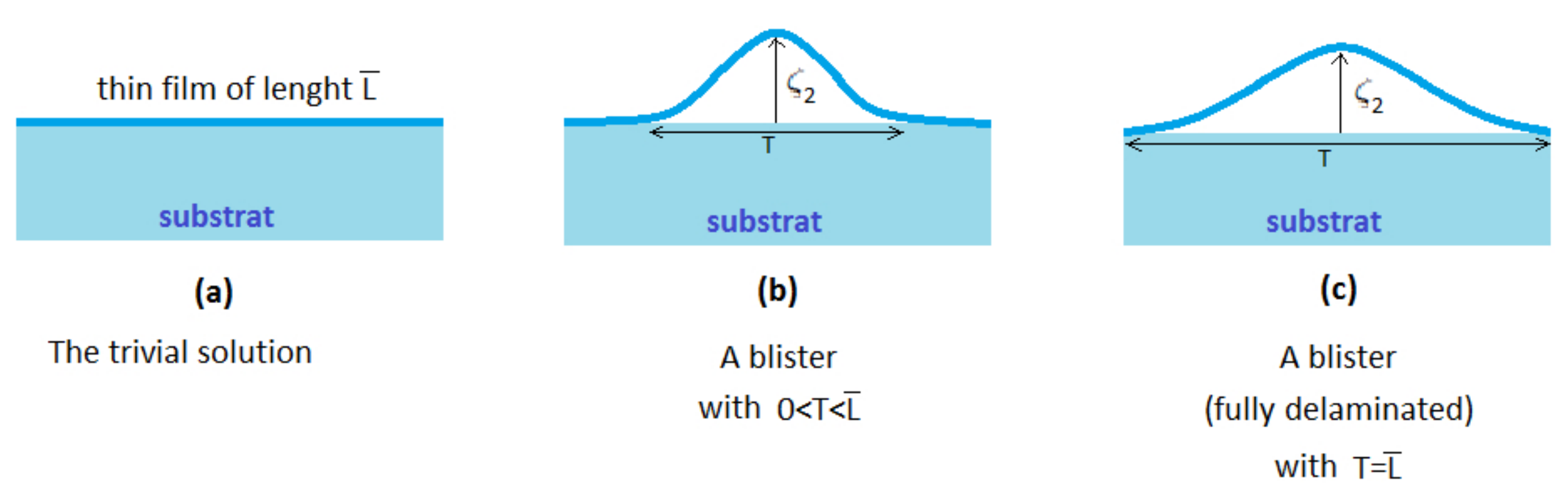}
\caption{Different types of solutions of problem
(\ref{min-E})}\label{blister}
\end{center}
\end{figure}

The thin films are often obtained by evaporation on a substrate.
When the coefficient of thermal expansion of the substrate is higher
than that of the film, cooling to ambient temperature leads to a
compressive residual stress in the film. If compression is
sufficient, the film tends to buckle, separating from the substrate.
It is said that the film delaminates (see Figure \ref{blister}).\\


\maketitle

\noindent An oversimplified one dimensional model which describes this phenomena is
given by the minimization of the following energy (of F\"oppel-von K\'arm\'an type)
\begin{eqnarray}\label{energyE}
E(\zeta_1,\zeta_2)&:=&\int_{\Omega}\gamma\mathbf{1}_{\{\zeta_2>0\}}+4\alpha\left(\zeta_1'+\frac12(\zeta_2')^2\right)^2+\frac{4\alpha}3\zeta_2''^2-2\overline{\theta}\zeta_2'^2
\quad \mbox{ with } \quad \gamma=1,
\end{eqnarray}
with
$$\Omega=\mathbb{R}/\overline{L}\mathbb{Z}=[-\overline{L}/2,\overline{L}/2[_{per},$$
and where $\overline{L}$ is the length of the film, $\alpha>0$
represents its elasticity coefficient and $\overline{\theta}>0$ is
the thermal parameter. Here the parameter $\gamma$ measures the cost
of delamination and is similar to the formulation of fracture with
Griffith criterion (see for instance Francfort, Marigo \cite{F-M}, Griffith \cite{G}, Larsen \cite{L}). For $\gamma=0$, this model was formally derived from 3D elasticity in the asymptotics of thin films in \cite{D} by El Doussouki and the last author, see also \cite{M}. For
simplicity, we normalize this parameter $\gamma$ to be equal to $1$
in the whole paper (this normalization can always be absorbed in a
redefinition of $E$, $\alpha$ and $\overline{\theta}$ by
rescaling). 
The quantity $\zeta_2(x)$ denotes the vertical displacement and is
assumed to be nonnegative (the film is above the substrate) and
$\zeta_1(x)$ is the horizontal one with $x\in\Omega$, where the
periodicity is assumed to simplify the analysis (see also Remark
\ref{rem-boundary-cond} for other boundary conditions describing a
clamped film). We introduce the following space
\begin{equation}\label{Y}
Y:=H^1\left(\Omega\right)\times\{ \zeta_2\in H^2\left(\Omega\right),
\zeta_2\geq 0\}.
\end{equation}

\noindent The solution of our model is given by solving the
following problem
\begin{equation}\label{min-E}
\min_{(\zeta_1,\zeta_2)\in Y}E(\zeta_1,\zeta_2).
\end{equation}
\begin{definition}\label{def-blister}(\textbf{Blisters})\\
We call a "blister" any global minimizer of the energy $E$ defined
in (\ref{energyE}) which is non trivial i.e.
$(\zeta_1,\zeta_2)\not\equiv (0,0)$.
\end{definition}

\noindent This paper elaborates the delamination of compressed thin films.
Under appropriate conditions, blisters may appear. We give a
complete description of global minimizers in terms of the parameters
of the problem.

\newpage

\subsection{Main results}\label{sub-sect-main-results}
\begin{theorem}\label{theo-blisters}(\textbf{Existence of global minimizers})\\
There exists a (global) minimizer $\zeta=(\zeta_1,\zeta_2)\in Y$ of
the energy $E$ introduced in (\ref{energyE}).
\end{theorem}

\noindent In order to study minimizers of $E$, it is useful to consider the
following auxiliary minimizing problem
\begin{equation}\label{min-f}
\min_{X\in\mathcal{D}}f(X),
\end{equation}
where
\begin{equation}\label{energyf}
f(X):=\left\{\begin{array}{ll}
(\theta-X)^{-1/2}-LX^2 \quad &\mbox{ if } \quad 0<X<\theta; \\
\\
0 &\mbox{ if } \quad X=0;\\
\end{array}\right.
\end{equation}
with rescaled versions of the thermal parameter $\overline{\theta}$
and of the length $\overline{L}$
\begin{equation}\label{a-b}
\theta:=\frac{\overline{\theta}}{\alpha}, \quad \mbox{and} \quad
L:=\frac{1}{2\pi}\sqrt{\frac{3}{2}}\alpha \overline{L},
\end{equation}
where $\alpha>0$ is from now on fixed in the model and $\mathcal{D}$
is the interval given by
\begin{equation}\label{D}
\mathcal{D}:=\left[0,\widetilde{\theta}^+\right],
\quad \mbox{ with } \quad \widetilde{\theta}^+=\max\left\{\widetilde{\theta},0\right\} \quad \mbox{and} \quad
\widetilde{\theta}:=\theta-\frac{\alpha^2}{L^2}.
\end{equation}
 Indeed the following theorem shows that the minimizing
problem (\ref{min-E}) is equivalent to the study of the
auxilary problem (\ref{min-f}).
\begin{theorem}\label{theo-description}(\textbf{Description of global minimizers of $E$})\\
{\bf i) \textit{(Implication)}}\\
For any global minimizer $\zeta=(\zeta_1,\zeta_2)$ of the
energy $E$, there exists at least a minimizer $K\in \mathcal{D}$ of
problem (\ref{min-f}) such that the following holds: there exists
$T\in [0,\overline{L}]$ such that (up to addition of constants and
translation of $(\zeta_1,\zeta_2)$), this minimizer $\zeta$ can be
written as follows 
\begin{equation}\label{zeta1-zeta2}
\left\{\begin{array}{ll} \zeta_1(x)&=\left\{\begin{array}{ll}
\displaystyle{\frac{K}{2}\left(x+\frac{\overline{L}}{2}\right)} \quad &\mbox{ in } \quad [-\overline{L}/2,-T/2]; \\
\\
\displaystyle{\frac{K\overline{L}}{8\pi}\sin(2\beta x)+\frac{K}{2}\left(1-\frac{\overline{L}}{T}\right)x} &\mbox{ in } \quad (-T/2,T/2);\\
\\
\displaystyle{\frac{K}{2}\left(x-\frac{\overline{L}}{2}\right)}
\quad &\mbox{ in } \quad [T/2,\overline{L}/2],
\end{array}\right.\\
\\
\zeta_2(x)&=\left\{\begin{array}{ll}
\displaystyle{A(\cos(\beta x)+1)} &\mbox{ ~~~~~~~~~~~~~~~~~~~in } \quad [-T/2,T/2]; \\
\\
0 \quad &\mbox{ ~~~~~~~~~~~~~~~~~~~elsewhere, }\\
\end{array}\right.\\
\end{array}
\right.
\end{equation}
where $\beta$, $A$ and $T$ are given by
\begin{equation}\label{beta-T-A}
\begin{array}{ll}
\beta &:=\displaystyle{\sqrt{\frac{3(\overline{\theta}-\alpha K)}{2\alpha}}}; \quad
A:=\displaystyle{\sqrt{\frac{K\overline{L}}{\pi \beta}}}; \quad
T:=\left\{\begin{array}{ll}
\displaystyle{\frac{2 \pi}{\beta}} \quad &\mbox{ if } \quad K>0; \\
\\
0 \quad &\mbox{ if } \quad K=0.\\
\end{array}\right.\\
\end{array}
\end{equation}
More generally, for any $K\in\mathcal{D}$ and any functions $(\zeta_1,\zeta_2)$ given in (\ref{zeta1-zeta2})-(\ref{beta-T-A}), we have
\begin{equation}\label{equ-E-f}
E(\zeta_1,\zeta_2)=2\pi\sqrt{\frac{2}{3}}f(K),
\end{equation}
and for $K\in\mathcal{D}$
\begin{equation}\label{T-K}
\left\{\begin{array}{lcl}
T=\overline{L} &\Leftrightarrow &K=\widetilde{\theta}; \\
\\
&\mbox{and}\\
\\
T<\overline{L} &\Leftrightarrow &K<\widetilde{\theta}. \\
\end{array}\right.
\end{equation}
{\bf ii) \textit{(Reciprocal)}}\\ 
If $K\in\mathcal{D}$ is a minimizer of problem
(\ref{min-f}), then the function $\zeta=(\zeta_1,\zeta_2)$ given in
(\ref{zeta1-zeta2})-(\ref{beta-T-A}) is a global minimizer of $E$
on $Y$.
\end{theorem}

\noindent Notice that $\overline{\theta}-\alpha K>0$ because
$K\in\mathcal{D}$. Moreover, when $K=0$ then $A=T=0$ which implies
that $\zeta_1=\zeta_2=0$. Thus with our definition, $T$ can be
interpreted as the length of the support of $\zeta_2$. Theorem
\ref{theo-description} identifies three types of global minimizers.
For $K=0$, we get the trivial solution (Figure \ref{blister}, (a)).
For $K\in(0, \widetilde{\theta})$, then $0<T<\overline{L}$ and we
get the blister solution (Figure \ref{blister}, (b)). Finally, for
$K=\widetilde{\theta}$, then $T=\overline{L}$ and we get the
fully delaminated blister (Figure \ref{blister}, (c)). We still use the name "blister" for the mathematical solution even if physically the film is completely delaminated. Note that our blister solution (Figure \ref{blister}, (b)) can be roughly speaking seen as the cross section of blisters with the shape of fingers (see for instance experiments in Figure 8.1 in \cite{V-W}).

\begin{figure}[htb]
\begin{center}
\includegraphics[width=0.97\textwidth]{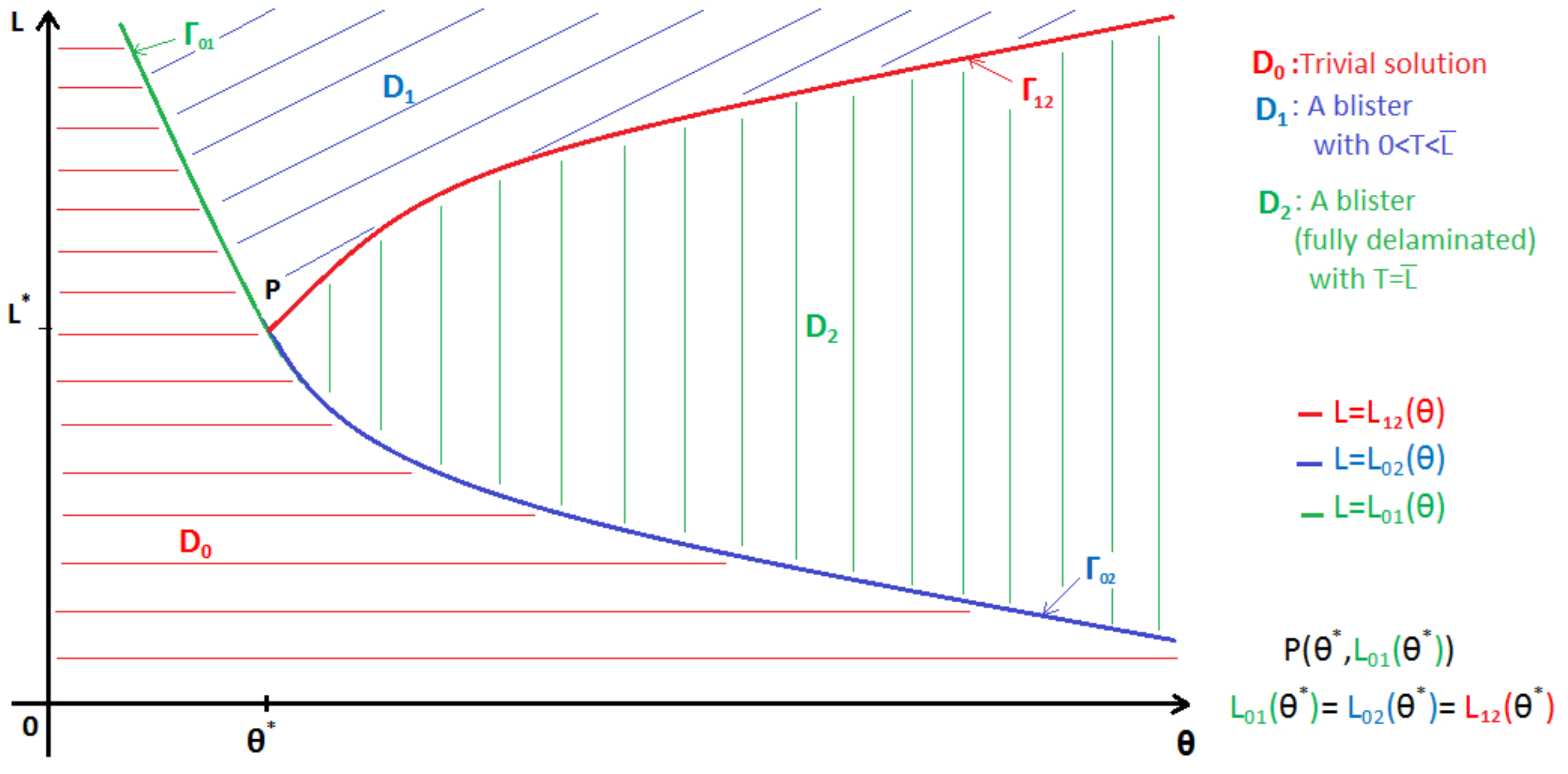}
\caption{Schematic phase diagram for parameters $(\theta,L)$}\label{fig-domain}
\end{center}
\end{figure}

\begin{remark}\label{rem-boundary-cond}(\textbf{Clamped boundary conditions})\\
Recall that the periodic boundary conditions are included in the set
$Y$ defined in (\ref{Y}). We now introduce another set of functions
satisfying clamped boundary conditions
$$\widetilde{Y}:=H^1_0(-\overline{L}/2,\overline{L}/2)\times\{\zeta_2\in H^2_0(-\overline{L}/2,\overline{L}/2), \zeta_2\geq 0\}.$$
Then
$$\inf_{\widetilde{Y}}E\geq \inf_{Y}E,$$
because any $y\in \widetilde{Y}$ can be seen as an element of $Y$
when it is extended by periodicity. Moreover, any global minimizer
of $E$ on $Y$ is given (up to addition of constants and translation
of $(\zeta_1,\zeta_2)$) by the solution written in
(\ref{zeta1-zeta2}) which satisfies $(\zeta_1,\zeta_2)\in
\widetilde{Y}$. Therefore,
$$\inf_{\widetilde{Y}}E= \inf_{Y}E,$$
and then in this paper we also solved the minimization problem of
$E$ on $\widetilde{Y}$.
\end{remark}
\noindent To classify the solutions obtained in Theorem
\ref{theo-description}, we have to define the following functions in
order to introduce some domains $D_0$, $D_1$ and  $D_2$ of
parameters $(\theta,L)$. Figure \ref{fig-domain} describes those
domains (still for arbitrary fixed value $\alpha$). We will show
that trivial solutions correspond to $D_0$, blister solutions to
$D_1$ and fully delaminated blister to $D_2$. For this purpose, we
introduce
\begin{numcases} \strut
\displaystyle{\theta^*:=\frac{5}{4}\alpha^{-1/2}};\label{a1}\\
\displaystyle{L_{01}(\theta):=\frac{5^{5/2}}{16}\theta^{-5/2}} ~~~~~~~~~~~~~~~~~~~~~~~~~~~~~\mbox{ for } 0<\theta\leq \theta^*; \label{b+}\\
\displaystyle{L_{02}(\theta):=\frac{\alpha^{5/4}}{(\sqrt{\alpha}\theta-1)^{1/2}}}~
~~~~~~~~~~~~~~~~~~~~~~~ \mbox{ for } \theta\geq \theta^*>\alpha^{-1/2};\label{F1}\\
\displaystyle{L_{12}(\theta):=\left(2\alpha^3\theta+2\alpha^2\sqrt{\alpha(\alpha\theta^2-1)}\right)^{1/2}} ~~\mbox{ for } \theta\geq \theta^*>\alpha^{-1/2}.\label{bx}
\end{numcases}

\begin{definition}(\textbf{Domains $D_0, D_1$ and $D_2$})\label{def-domain}\\
Let us now introduce the following sets of $(\theta,L)\in \left(0,+\infty\right)^2$:
\begin{numcases} \strut
D_0:=\left\{
\begin{array}{lll}
(\theta,L), &L< L_{01}(\theta) &\mbox{if } ~0<\theta\leq \theta^*\\
\\
&L<L_{02}(\theta) &\mbox{if } ~\theta>\theta^* \\
\end{array}
\right\};\label{D3}\\
\nonumber\\
D_1:=\left\{
\begin{array}{lll}
(\theta,L), &L> L_{01}(\theta) &\mbox{if } ~0<\theta\leq \theta^*\\
\\
&L>L_{12}(\theta) &\mbox{if } ~\theta>\theta^* \\
\end{array}
\right\};\label{D1}\\
\nonumber\\
D_2:=\left\{(\theta,L), \theta>\theta^* \mbox{ and } 
L_{02}(\theta)<L< L_{12}(\theta)\right\}.\label{D2}
\end{numcases}
We denote by
\begin{numcases} \strut
\Gamma_{01}:=\left\{(\theta,L),
0<\theta<\theta^* \mbox{ and } L=L_{01}(\theta)\right\};\\
\nonumber\\
\Gamma_{02}:=\left\{(\theta,L),
\theta> \theta^* \mbox{ and } L=L_{02}(\theta)\right\};\\
\nonumber\\
\Gamma_{12}:=\left\{(\theta,L),
\theta>\theta^* \mbox{ and }
L=L_{12}(\theta)\right\};\\
\nonumber\\
P=(\theta^*, L_{01}(\theta^*)).
\end{numcases}
\end{definition}
\begin{remark}\label{rem-partition}(\textbf{A partition of the domains})\\
We have the following disjoint decomposition
$$ \left(0,+\infty\right)^2=D_0\cup D_1\cup D_2 \cup \Gamma_{21}\cup\Gamma_{01}\cup\Gamma_{02}\cup\{P\}.$$ 
Moreover, the following properties hold true
\begin{equation}\label{curve-property}
\left\{\begin{array}{ll}
L_{12}'(\theta)>0 \mbox{ and } L_{02}'(\theta)<0, \quad &\mbox{ for } \theta>\theta^*;\\
\\
L_{01}'(\theta)<0, \quad &\mbox{ for } 0<\theta<\theta^*;\\
\\
L^*:=L_{12}(\theta^*)=L_{01}(\theta^*)=L_{02}(\theta^*);&
\end{array}\right.
\end{equation}
where $\theta^*$ is defined in (\ref{a1}).
\end{remark}

\noindent The proof of Remark \ref{rem-partition} is done by a simple computation.

\newpage

\begin{theorem}\label{theo-class-blisters}(\textbf{Classification of global minimizers of $E$})\\
{\bf i)} For $(\theta,L)\in D_0$, the unique global minimizer of the energy $E$ introduced in (\ref{energyE}) is the trivial solution $(\zeta_1,\zeta_2)=(0,0)$.\\
{\bf ii)} For $(\theta,L)\in D_1\cup D_2\cup \Gamma_{12}$, there is a unique blister $\zeta=(\zeta_1,\zeta_2)\in Y$ (see Definition \ref{def-blister}) minimizing the energy $E$. Moreover, the
component $\zeta_2$ has a support of length $T$ which is defined in
(\ref{beta-T-A}) and
\begin{equation}\label{T-domains}
\left\{\begin{array}{ll} T<\overline{L} & \quad \mbox{if } \quad
(\theta,L)\in
D_1 ;\\
\\
T=\overline{L} & \quad \mbox{if } \quad (\theta,L)\in D_2\cup \Gamma_{12}.\\
\end{array}\right.
\end{equation}
{\bf iii)} For $(\theta,L)\in \Gamma_{01}\cup \Gamma_{02}\cup \{P\}$,
the energy $E$ has exactly two global minimizers: the trivial
solution $\zeta=(\zeta_1,\zeta_2)=(0,0)$ and a blister
$\zeta=(\zeta_1,\zeta_2)\in Y$ given in (\ref{zeta1-zeta2}) with
\begin{equation}\label{T-boundary}
\left\{\begin{array}{ll} T<\overline{L} & \quad \mbox{if } \quad
(\theta,L)\in
\Gamma_{01};\\
\\
T=\overline{L} & \quad \mbox{if } \quad (\theta,L)\in \Gamma_{02}\cup\{P\}.\\
\end{array}\right.
\end{equation}

\end{theorem}

\begin{proposition}\label{prop-property}(\textbf{Blister's properties in $D_1\cup D_2\cup\Gamma_{12}$})\\
For $(\theta,L)\in D_1\cup D_2\cup\Gamma_{12}$, there exists a
unique $K\in \mathcal{D}$ (depending on $(\theta,L)$) minimizing
problem (\ref{min-f}). Recalling (\ref{a-b}), we consider $T$ and
$A$ given in (\ref{beta-T-A}). \\
{\bf i) Monotonicity }\\ 
First, $T$ and $A$ are continuous in
$(\theta,L)$ on $D_1\cup D_2\cup\Gamma_{12}$ and satisfy the
following properties
$$\frac{\partial T}{\partial \theta}, \frac{\partial T}{\partial L}\geq 0 \quad \mbox{and} \quad \frac{\partial A}{\partial \theta}, \frac{\partial A}{\partial L}\geq 0.$$
In particular,
$$T=\overline{L} \quad  \mbox{ on } \quad D_2\cup\Gamma_{12}.$$
{\bf ii) "Smallest" blister solutions }\\ 
We have  
\begin{equation}\label{T-star}
\inf_{(\theta, L)\in D_1}T=T^{*}:=4\pi \sqrt{\frac{2}{3}}\alpha^{1/4},
\end{equation}
and
\begin{equation}\label{A-star}
\inf_{(\theta, L)\in D_1}A=A^{*}:=\frac{4}{\sqrt{3}}.
\end{equation}
\end{proposition}


\begin{remark}(\textbf{Prediction for the smallest blisters; not fully delaminated case})\\
For any $(\theta, L)\in D_1$, we have a unique blister $(\zeta_1,\zeta_2)\in Y$ minimizing the energy $E$. According to Proposition \ref{prop-property}, the second component $\zeta_2$ has a support of length $T>T^*$ with $T<\overline{L}$. This shows that $T^{*}$ can be interpreted as the infinimum of the width of blisters whose length support is strictly less than the length of the film. Similarly, we can also interpret the amplitude $A^{*}$ as the minimal amplitude of the blisters. 
\end{remark}


\begin{remark}\label{rem-L-infty}(\textbf{Relatively small blisters for large $L$})\\
For $(\theta,L)\in D_1$, it is possible to check, as $L$ tends to
infinity, that $T$ and $A$ have a behavior like $L^{1/3}$ and
$L^{2/3}$ respectively. In particular, for $\theta$ fixed and for
large enough films, the size of the blisters is much smaller than
the size of the film.
\end{remark}

\begin{remark}(\textbf{Phase diagram for local minimizers})\\
For local minimizers (with small perturbation of the support with $T< \overline{L}$), we expect to have similar phase diagram for $p$ blisters (only local minimizers) of the same width $T$ with now $\overline{L}$ replaced by $\overline{L}/p$, $L$ replaced by $L/p$ and $T\leq \overline{L}/p$
(see Remark \ref{rem-p-blister}).
\end{remark}

\subsection{Brief review of the literature}
\noindent Buckling delamination blisters have commonly been studied for a long time. 
In \cite{G-O}, Gioia and Ortiz give an overview of experiments, propose and study mathematically variational models of blisters, among other things motivated by the description of telephone-cord morphology. See also \cite{A2} where such a telephone-cord instability is studied.\\
Experimentally and theoretically in \cite{V-B-B-R-R}, the authors study blisters which have the one dimensional symmetry. Their results seem coherent with ours, even if the problem and the modeling are not exactly the same. We also refer the reader to \cite{B-K} and the references therein for recent developments on the analysis and modeling of blisters. In this nice work, the authors consider a F\"oppel-von K\'arm\'an model for the film with a special bonding energy with the substrate. For this variational model, they study several regimes for the energy. This is also interesting to mention the work \cite{B-B-M-M} where the authors derive rigorously a variational similar model of thin films bonded to a substrate when the thickness of the film goes to zero. Their limit energy contains in particular a bonding term which is similar to our term with $\gamma$ in the energy (\ref{energyE}).

\subsection{Organization of the paper}
\noindent The organization of the paper is as follows. In Section
\ref{sec-proof-theo}, we prove Theorems \ref{theo-blisters} and
\ref{theo-description} i) on the existence and the description of global minimizers.
Section \ref{sec-min-aux-func} is dedicated to the detailed classification of global minimizers and their qualitative properties. There we prove Theorem \ref{theo-class-blisters}, Theorem \ref{theo-description} ii) and Proposition \ref{prop-property}. To this end, we divided this section into three parts. In the first one, we present some results which will be useful to prove Theorem \ref{theo-class-blisters}. The second subsection is devoted to prove Theorem \ref{theo-class-blisters}.  We end up Section \ref{sec-min-aux-func} by the proofs of Theorem \ref{theo-description} ii) and Proposition \ref{prop-property}.

\section{Proofs of Theorems \ref{theo-blisters} and \ref{theo-description} i)}\label{sec-proof-theo}
This section is divided into two parts: the first one is devoted to prove Theorem \ref{theo-blisters} and the second is dedicated to the proof of Theorem  \ref{theo-description} i).
\subsection{Existence of global minimizers}

\noindent\textbf{Proof of Theorem \ref{theo-blisters}} \\
The proof of Theorem \ref{theo-blisters} is very classical. 
By considering a minimizing sequence
$\zeta^k=(\zeta_1^k,\zeta_2^k)\in Y$ such that
$E(\zeta^k)\underset{k\to\infty}{\longrightarrow}I=\underset{\zeta\in Y}{\inf}E(\zeta)$, and using
 Young's inequality, we get
\begin{equation}\label{young-ineq}
2\overline{\theta}\int_{\Omega}(\zeta_2^{k'})^2=4\overline{\theta}\int_{\Omega}\left(\zeta_1^{k'}+\frac{1}{2}(\zeta_2^{k'})^2\right)\leq
\frac{2\overline{\theta}^2\overline{L}}{\alpha}+2\alpha\int_{\Omega}\left(\zeta_1^{k'}+\frac{1}{2}(\zeta_2^{k'})^2\right)^2.
\end{equation}
Thus we can bound the energy $E(\zeta^k)$. We will skip the steps of the proof since the result can be obtained in a classical way (see also \cite{M} and \cite{C}).

\hfill $\Box$

\subsection{Description of global minimizers of $E$}
We first start this subsection by the following lemma which will be used
to prove Theorem \ref{theo-description} i).
\begin{lemma}\label{lem-pde}(\textbf{Classification of solutions $\mathbf{\zeta_2}$})\\
Let $\zeta_2\in \{f\in H^2(\Omega), f\geq 0\}$. Consider the
following ordinary differential equation
\begin{equation}\label{edp-zeta3}
\left\{\begin{array}{ll}
\displaystyle{\zeta_2^{(4)}+\frac{3(\overline{\theta}-\alpha
K)}{2\alpha}\zeta_2''=0} \quad &\mbox{ on } \quad
\omega_0=(x_0,y_0);\\
\\
\zeta_2>0 \quad &\mbox{ on } \quad \omega_0;\\
\\
\zeta_2=0 \quad &\mbox{ on } \quad \partial\omega_0;\\
\end{array}\right.
\end{equation}
where $x_0<y_0$.\\
If $\overline{\theta}-\alpha K\leq 0$, then there is no solution of
(\ref{edp-zeta3}).\\
If $\overline{\theta}-\alpha K>0$, then up to translate $\zeta_2$, we have
\begin{equation}\label{zeta-T-beta}
\left\{\begin{array}{ll}
\displaystyle{x_0=-\frac{T}{2}} \quad \mbox{ and} \quad \displaystyle{y_0=\frac{T}{2}};\\
\\
\zeta_2(x)=A_0(\cos(\beta x)+1) \quad \mbox{ on } \quad \omega_0=(x_0,y_0);\\
\\
\beta=\displaystyle{\sqrt{\frac{3(\overline{\theta}-\alpha
K)}{2\alpha}}=\frac{2\pi}{T}};\\
\end{array}\right.
\end{equation}
where $A_0>0$ is a constant.
\end{lemma}

\noindent {\bf Proof of Lemma \ref{lem-pde}}\\
Since $\zeta_2\in H^2(\mathbb{R}/\overline{L}\mathbb{Z})$, then
$\zeta_2\in C^1(\mathbb{R}/\overline{L}\mathbb{Z})$. Moreover
$\zeta_2\geq 0$, which implies that
\begin{equation}\label{boudary-cond}
\zeta_2(x_0)=\zeta_2'(x_0)=0=\zeta_2(y_0)=\zeta_2'(y_0).
\end{equation}
We can write $\zeta_2$ as  
$\zeta_2(x)=\zeta_2^S(x)+\zeta_2^A(x)  \mbox{ on } \omega_0=(x_0,y_0)$,
where $\zeta_2^S$ is
the symmetric part of $\zeta_2$ and $\zeta_2^A$ is its anti-symmetric part. In particular $\zeta_2^S$ verifies the following conditions
\begin{equation}\label{sym-conditions}
\zeta_2^S>0 \quad \mbox{ on } \omega_0 \quad \mbox{ and } \quad
\zeta_2^S=(\zeta_2^S)'=0 \quad \mbox{ on } \partial \omega_0.
\end{equation}
We skip the details of the proof which is a routine exercise.

\hfill $\Box$

\noindent{\textbf{Proof of Theorem \ref{theo-description} i)}}\\
Let $(\zeta_1,\zeta_2)\in  Y$ be a minimizer of $E$.\\
\textbf{Step 1: Differentiating $E$ with respect to $\zeta_1$}\\
Differentiating $E$ with respect to $\zeta_1$ leads to the following
Euler-Lagrange equation:
$$\left(\zeta_1'+\frac12(\zeta_2')^2\right)'=0 \quad \mbox{on} \quad \Omega,$$
i.e.
\begin{equation}\label{rel-zeta3-zeta1}
\zeta_1'+\frac12(\zeta_2')^2=\frac K2 \quad \mbox{ where } \quad K=\frac{1}{\overline{L}}\int_{\Omega}(\zeta_2')^2\geq 0.
\end{equation}
Therefore the total energy becomes
\begin{eqnarray*}\label{energy1}
E(\zeta_1,\zeta_2)=\overline{E}(\zeta_2)&:=&\int_{\Omega}\mathbf{1}_{\{\zeta_2>0\}}+\frac{\alpha}{\overline{L}}{\left(\int_{\Omega}(\zeta_2')^2\right)}^2
+\frac{4\alpha}3\int_{\Omega}(\zeta_2'')^2 -2\overline{\theta}
\int_{\Omega}(\zeta_2')^2.
\end{eqnarray*}
If $K=0$, then $\zeta_2\equiv 0$ and thus $\zeta_1\equiv \mbox{const}$ on
$\Omega$, and up to subtract a constant to $\zeta_1$, we can assume that $\zeta_1\equiv 0$.\\
If $K>0$, then $\zeta_2\not\equiv 0$ and we proceed as follows.\\
\textbf{Step 2: Differentiating $\overline{E}$ with respect to $\zeta_2$}\\
Differentiating $\overline{E}$ with respect to $\zeta_2$, yields the
following Euler-Lagrange equation
$$
\zeta_2^{(4)}+\frac{3(\overline{\theta}-\alpha
K)}{2\alpha}\zeta_2''=0 \quad \mbox{ on } \quad \{\zeta_2>0\}.
$$
Up to add a constant to $\zeta_2$, we can assume that
$\underset{\Omega}{\inf}\zeta_2=0$. Therefore there exists
$x_0\in\Omega$ such that $\zeta_2(x_0)=0$. Up to translation, we
choose $x_0=-\overline{L}/2$. Then, we deduce that
$$\{\zeta_2>0\}=\underset{i\in J}{\cup}\omega_i, \quad \mbox{ for } \quad \omega_i=(x_i,y_i),$$
where $J$ is a set at most countable and such that
$\omega_i\cap\omega_j=\emptyset$ for $i\neq j$. Applying Lemma
\ref{lem-pde} to each $\omega_i$, we conclude that
$\overline{\theta}-\alpha K>0$,
$$y_i-x_i=T=2\pi\sqrt{\frac{2\alpha}{3(\overline{\theta}-\alpha
K)}}$$ and up to translation, the solution $\zeta_2$ is given by
(\ref{zeta-T-beta}) on each $\omega_i$ with the amplitude $A_0$ replaced by $A_i$. Now, we deduce that
$\mbox{card}(J)=p<+\infty$ with $p\geq 1$ satisfying $p T\leq
\overline{L}$. Since the $\omega_i$ are disjoint, we get
\begin{equation}\label{T-K-A}
K\overline{L}=\int_{\Omega}(\zeta_2')^2=\frac{\beta^2T}{2}\sum_{i=1}^pA_i^2.
\end{equation}
Hence using (\ref{zeta-T-beta}), we get
\begin{eqnarray}\label{energy2}
\overline{E}(\zeta_2)=\overline{\overline{E}}(K,p)&:=&2p\pi\sqrt\frac{2\alpha}{3(\overline{\theta}-\alpha
K)}-\alpha K^2 \overline{L}
 \end{eqnarray}
with the conditions
\begin{equation}\label{beta-p}
1\leq p\leq \frac{\overline{L}}{T} \quad \mbox{ and } \quad
T=\frac{2\pi}{\beta}.
\end{equation}
Then we minimize the energy with respect to $p$ and we get the results.

\hfill $\Box$

\begin{remark}\label{rem-p-blister}(\textbf{Local minimizers})\\
For local minimizers of $E$, we may have $p$ blisters (all separated by any positive distance) with the same width $T$ and with amplitude $A_i$ satisfying (\ref{T-K-A}). For $p\geq 1$ given, we can also optimize $K$ in $\overline{\overline{E}}(K,p)$ which should correspond to local minimizers of $E$ (restricted to small perturbations of the support with $T<\overline{L}$) with $\overline{L}$ replaced by $\overline{L}/p$, $L$ replaced by $L/p$ and $T\leq \overline{L}/p$.
\end{remark}

\section{Proofs of Theorem \ref{theo-class-blisters}, Theorem \ref{theo-description} ii) and Proposition \ref{prop-property}}\label{sec-min-aux-func}

Our aim is to prove Theorem \ref{theo-class-blisters}, Theorem \ref{theo-description} ii) and Proposition \ref{prop-property}. For this purpose, this section is divided into several parts. In the first one, we give some tools which will be useful to prove Theorem \ref{theo-class-blisters}. The second subsection is dedicated to the proof of Theorem \ref{theo-class-blisters}. Finally, we prove Theorem \ref{theo-description} ii) and Proposition \ref{prop-property} in the last subsection.

\subsection{Preliminaries}

First, we are interested in the following auxiliary
minimization problem
\begin{equation}\label{aux-min-f}
\min_{X\in \mathcal{D}}f(X)=f(K),
\end{equation}
where $f$ and $\mathcal{D}$ are defined respectively in
(\ref{energyf}) and (\ref{D}).
\noindent Recall that for $(\theta,L)\in(0,\infty)^2$, we have that
$\mathcal{D}\underset{\neq}{\subset}[0,\theta)$.
In order to determine the minimum of the function $f$ on $[0,\theta)$, we have to introduce the quantity 
\begin{equation}\label{b-star-0}
L_d=L_d(\theta):=\frac{25}{24}\sqrt{\frac{5}{3}}\theta^{-5/2}, \quad \mbox{ for } \theta>0.
\end{equation}

\begin{proposition}\label{prop-X_m}(\textbf{Minimizing $f$ on $[0,\theta)$ and its consequences})\\
We consider $f$ defined in (\ref{energyf}) and $L_d$ introduced in (\ref{b-star-0}).\\
{\bf \textit{i)}} If $L\leq L_d$, then $f$ is increasing on $(0,\theta)$ and
$$\underset{\mathcal{D}}{\operatorname{argmin}}f= \{0\}.$$
In this case, we set artificially $X_m:=2\theta/5$.\\
{\bf \textit{ii})}  If $L> L_d$, then there exist $X_M$ and $X_m$ such that
$0<X_M<2\theta/5<X_m<\theta$ and
$$
\left\{\begin{array}{ll}
f'\geq 0 \quad &\mbox{ on } \quad (0,X_M]\cup[X_m,\theta); \\
\\
f'<0  \quad &\mbox{ on } \quad (X_M,X_m);
\end{array}\right.
$$
and
\begin{equation}\label{argmin-f}
\underset{\mathcal{D}}{\operatorname{argmin}}f\subset \{0, X_m\}.
\end{equation}
{\bf \textit{iii})} Moreover for $L>L_d$, the function $X_m=X_m(\theta, L)$ is smooth and satisfies
\begin{equation}\label{deriv-K-theta}
\partial_{\theta}X_m=\frac{3/4(\theta-X_m)^{-5/2}}{3/4(\theta-X_m)^{-5/2}-2L}>1,
\end{equation}
and
\begin{equation}\label{deriv-K-L}
\partial_LX_m=\frac{2X_m}{3/4(\theta-X_m)^{-5/2}-2L}>0.
\end{equation}
\end{proposition}

\noindent{\bf Proof of Proposition  \ref{prop-X_m}}\\
{\bf Step 1: Proof of i) and ii)}\\ 
For  $0< X<\theta$, we have
$$f'(X)=g(X)-h(X) \quad \mbox{with} \quad  g(X):=\frac12(\theta-X)^{-3/2} \mbox{ and } h(X):=2LX.$$
We notice that $g$ is strictly convex on $[0,\theta)$, with $g(0)>0$ and $g'(0)>0$. Therefore there exists a unique $L=L_d>0$ such that the straight line $y=h(X)$ is tangent from below to the graph $y=g(X)$, at a point $X_d>0$. In particular, we have
$$\left\{\begin{array}{ll}
g(X_d)=h(X_d); \\
\\
g'(X_d)=h'(X_d).\\
\end{array}\right.$$
The unique solution of this system is $X_d=2\theta/5$ and the value of $L_d$ given in (\ref{b-star-0}).
Using the strict convexity of $g$ (and the fact that $g(\theta^-)=+\infty$), we deduce the variations of $f$ in cases $i)$ and $ii)$.
With the notations of case $ii)$ in Proposition \ref{prop-X_m}, $X_m$ is in particular uniquely characterized by
\begin{equation}\label{caract-X_m}
 f'(X_m)=0 \quad \mbox{with} \quad X_m\in \left(\frac{2\theta}{5},\theta\right),
\end{equation}
and then we get (\ref{argmin-f}).\\
{\bf Step 2: Proof of iii)}\\
In order to compute the derivative with respect to $(\theta, L)$, we write the dependance of $f$ on $(\theta, L)$ as: $f(X)=f(X,\theta, L)$. 
Using (\ref{caract-X_m}) we have
\begin{equation}\label{f''(K)}
\partial_{XX}^2f(X_m, \theta, L)=\frac{3}{4}(\theta-X_m)^{-5/2}-2L=\frac{L(5X_m-2\theta)}{\theta-X_m}>0.
\end{equation}
Using (\ref{f''(K)}), we have $\partial_{XX}f(X_m, \theta, L)\neq 0$. Then using the Implicit Function Theorem, we deduce that $X_m=X_m(\theta, L)$ is a smooth function. Now using the definition of $X_m=X_m(\theta, L)$, we get
\begin{equation}\label{derive=0}
\frac{d}{d\theta}(\partial_Xf(X_m,\theta, L))=\partial_{XX}^2f(X_m,\theta, L)\partial_{\theta}X_m+\partial_{X\theta}^2f(X_m,\theta, L)=0.
\end{equation}
Using (\ref{derive=0}) and (\ref{f''(K)}), we get (\ref{deriv-K-theta}). In a similar way, we get (\ref{deriv-K-L}).

\hfill $\Box$

\noindent  In what follows, we consider the minimizer
 of $f$ on the subinterval $\mathcal{D}$ of $[0,\theta)$
where we recall that $\mathcal{D}:=\left[0,\widetilde{\theta}^+\right]$ and $\widetilde{\theta}^+$ is defined in (\ref{D}). For this purpose, we introduce
\begin{equation}\label{X-bar}
\overline{X}:=\min\left\{\widetilde{\theta}^+,X_m\right\},
\end{equation}
where $X_m$ is the quantity introduced in Proposition
\ref{prop-X_m}. Then we have
\begin{equation}\label{min-D}
\underset{\mathcal{D}}{\operatorname{argmin}}f\subset \{0, \overline{X}\}.
\end{equation} 
For this reason, we have to study in particular the equalities
$$\left\{\begin{array}{lcl}
X_m &=& \widetilde{\theta}; \\
\\
f(0) &=& f(X_m);\\
\\
f(0) &=& f(\widetilde{\theta}).
\end{array}\right.$$

\noindent And then we need to consider the following functions
\begin{numcases} \strut
\displaystyle{L_d(\theta):=\frac{25}{24}\sqrt{\frac{5}{3}}\theta^{-5/2}}~~
~~~~~~~~~~~~~~~~~~~~~~~ \mbox{ for } \theta>0;\label{L-d}\\
\displaystyle{L_{01}(\theta):=\frac{5^{5/2}}{16}\theta^{-5/2}} ~~~~~~~~~~~~~~~~~~~~~~~~~~~\mbox{ for } \theta>0; \label{L01}
\end{numcases}
where $L_d$ and $L_{01}$ have already been introduced in (\ref{b-star-0}) and (\ref{b+}).\\

\begin{figure}[htb]
\begin{center}
\includegraphics[width=1\textwidth]{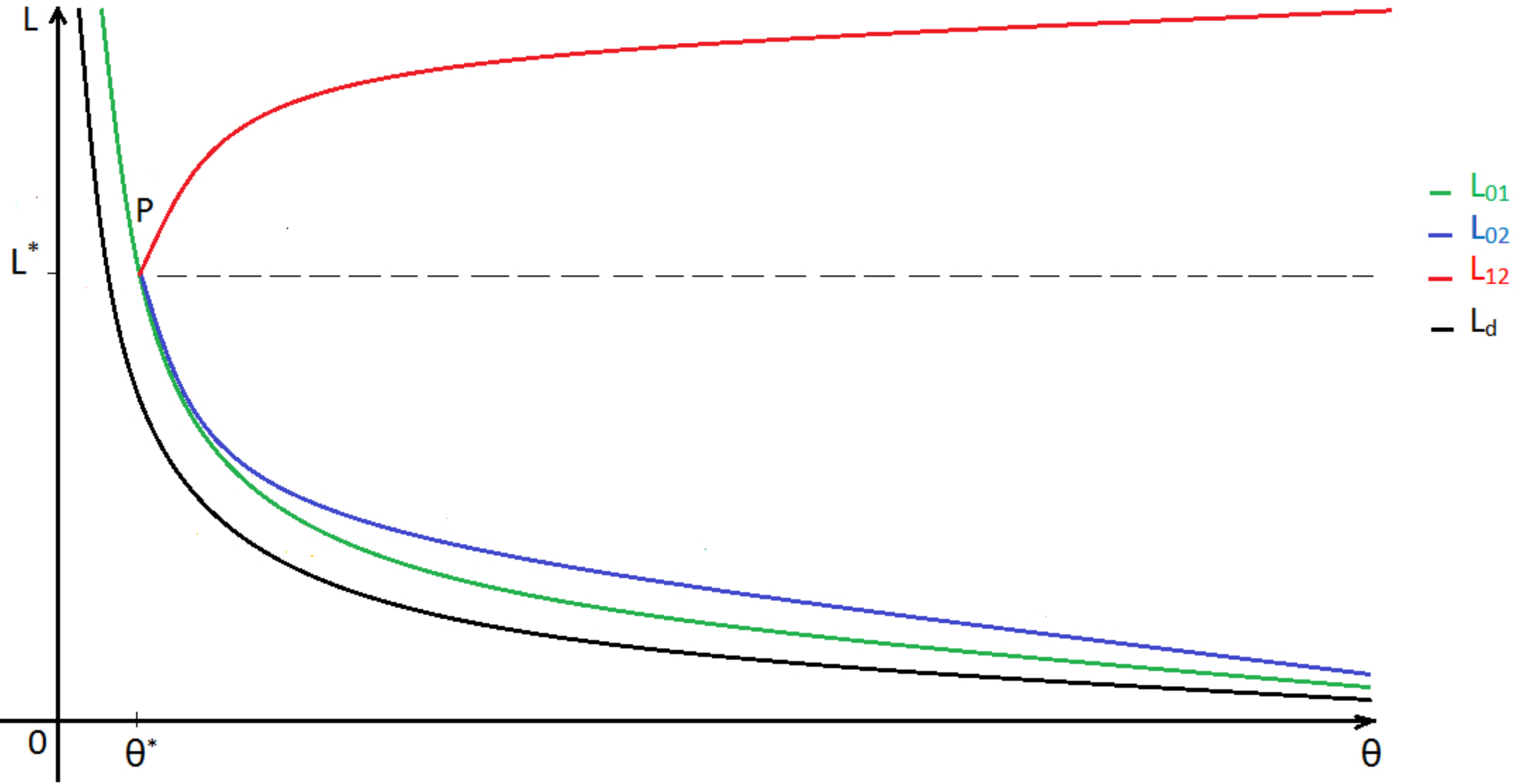}
\caption{Schematic phase diagram for parameters $(\theta,L)$}\label{fig-all-curves}
\end{center}
\end{figure}

\noindent First of all, we have to give some geometrical results concerning the position of such curves describing our domains. For an illustration of the following lemma, we refer the reader to Figure \ref{fig-all-curves}.

\begin{lemma}\label{rem-position}(\textbf{Positions of some curves})\\
We recall $\theta^*$ given in (\ref{a1}). The following results hold true:\\
{\bf i)} $L_{01}(\theta^*)=L_{02}(\theta^*)=L_{12}(\theta^*)$.\\
{\bf ii)} For all $\theta>\theta^*$, we  have
$L_{01}(\theta)< L_{02}(\theta)<L_{12}(\theta)$.\\
{\bf iii)} For all $\theta>0$, we have
$L_{01}(\theta)>L_d(\theta)$.

\end{lemma}
We skip the proof of Lemma \ref{rem-position} since it is easy to check the result by simple computations.

\newpage

\begin{lemma}\label{lem-Xm-theta-tilda}(\textbf{Properties of the function $X_m-\widetilde{\theta}$})\\
{\bf i)} For $\theta>0$ and $L>L_d(\theta)$, we have $\partial_{\theta}(X_m-\widetilde{\theta})>0$, where $X_m=X_m(\theta, L)$ and $\widetilde{\theta}=\widetilde{\theta}(\theta, L)$ are respectively introduced in Proposition \ref{prop-X_m} and (\ref{D}).\\
{\bf ii)} For $\theta>0$ and $L=L_{01}(\theta)$, we have $L>L_d(\theta)$ and
$$
\left\{\begin{array}{ll}
 X_m<\widetilde{\theta} \quad  &\mbox{ for } \quad 0<\theta< \theta^*; \\
\\
 X_m=\widetilde{\theta} \quad  &\mbox{ for } \quad \theta=\theta^*;\\
\\
X_m>\widetilde{\theta} \quad  &\mbox{ for } \quad \theta>\theta^*.
\end{array}\right.
$$
{\bf iii)} For $\theta\geq \theta^*$ and $L=L_{12}(\theta)$, we have $L> L_d(\theta)$ and $X_m=\widetilde{\theta}$.
\end{lemma}

\noindent {\bf Proof of Lemma \ref{lem-Xm-theta-tilda}}\\
{\bf Proof of i)} Let $\theta>0$ and $L> L_d(\theta)$. According to Proposition \ref{prop-X_m} ii), $f$ admits a non zero minimizer
$X_m\in
(2\theta/5,\theta)$.
Using (\ref{deriv-K-theta}) and the definition of $\widetilde{\theta}$ in (\ref{D}), we get
$$\partial_{\theta}(X_m-\widetilde{\theta})=\frac{3/4(\theta-X_m)^{-5/2}}{3/4(\theta-X_m)^{-5/2}-2L}-1=\frac{2L}{3/4(\theta-X_m)^{-5/2}-2L}>0.$$

\noindent {\bf Proof of ii)} Let $\theta>0$ and $L=L_{01}(\theta)$. Using Lemma \ref{rem-position}, we deduce that $L=L_{01}(\theta)>L_{d}(\theta)$ for all $\theta>0$. Then according to Proposition \ref{prop-X_m} ii), $f$ admits a non zero minimizer
$X_m\in
(2\theta/5,\theta)$.
Using (\ref{caract-X_m}) we have
$$\frac{1}{2}(\theta-X_m)^{-3/2}=2L_{01}(\theta)X_m \quad \mbox{ with } \quad \frac{2}{5}\theta< X_m<\theta.$$
 Let $Y_m:=X_m/\theta$,  then we have
 \begin{equation}\label{eq-Xm-bar}
 \frac{1}{2}(1-Y_m)^{-3/2}=\frac{5^{5/2}}{8} Y_m \quad \mbox{with}  \quad \frac{2}{5}< Y_m<1.
 \end{equation}
The uniqueness of $Y_m$ shows that $Y_m$ is a constant independent of $\theta$. It is easy to check that $Y_m=4/5$ is the solution of (\ref{eq-Xm-bar}). Then
\begin{equation}\label{K-K-bar}
X_m=\frac{4}{5}\theta  \quad \mbox{ for all } \theta>0 \mbox{ and } L=L_{01}(\theta).
\end{equation}
Using (\ref{K-K-bar}), we get that for $\theta>0$ and $L=L_{01}(\theta)$
$$X_m-\widetilde{\theta}=\frac{4\theta}{5}-\widetilde{\theta}(\theta, L_{01}(\theta))=\frac{\theta}{5}\left(\frac{4^4\alpha^2}{5^4}\theta^4-1\right),$$
which vanishes for $\theta=\theta^*$ and then we get the result.

\noindent {\bf Proof of iii)} Let $\theta\geq \theta^*$ and $L=L_{12}(\theta)$. Using Lemma \ref{rem-position}, we conclude that $L=L_{12}(\theta)\geq L_{01}(\theta) > L_d(\theta)$ for all $\theta\geq \theta^*$. Then using Proposition \ref{prop-X_m} ii), $f$ admits a non zero minimizer
$X_m\in
(2\theta/5,\theta)$. A direct computation of $f'(\widetilde{\theta})$ shows that
$$f'(\widetilde{\theta})=\frac{Q(\theta,L)}{2\alpha^3L} \quad \mbox{ with } \quad Q(\theta,L):=L^4-4\alpha^3\theta L^2+4\alpha^5.$$
It is easy to check that in particular for $\theta\geq \theta^*$ and $L=L_{12}(\theta)$, we have $Q(\theta, L_{12}(\theta))=0$ and then $f'(\widetilde{\theta})=0$.
Using Lemma \ref{lem-Xm-theta-tilda} ii), we know that 
\begin{equation}\label{property-P}
\widetilde{\theta}=X_m>\frac{2}{5}\theta \quad \mbox{ at the point } \quad (\theta, L)=(\theta^*, L^*)=P.
\end{equation}
Moreover for $\theta>\theta^*$ and $L=L_{12}(\theta)$, we have 
$$\frac{d}{d \theta}\left(\widetilde{\theta}(\theta, L_{12}(\theta))-\frac{2}{5}\theta\right)=\frac{3}{5}+2\frac{\alpha^2}{ L_{12}^3(\theta)} L'_{12}(\theta)>0.$$ 
Therefore we deduce that $\widetilde{\theta}>2\theta/5$ for $\theta\geq \theta^*$ and $L=L_{12}(\theta)$. Then we conclude that $X_m=\widetilde{\theta}$ for $\theta\geq \theta^*$ and $L=L_{12}(\theta)$.

\hfill $\Box$

\begin{lemma}\label{lem-f-theta-L}(\textbf{Properties of $f(X_m)$ and $f(\widetilde{\theta})$})\\
Let $f(X)=f(X,\theta,L)$ be given in (\ref{energyf}).\\
{\bf A) Properties of $\mathbf{\textit{f}(\textit{X}_m)}$}\\
For $\theta>0$ with $L> L_d(\theta)$ and $X_m=X_m(\theta, L)$ introduced in Proposition \ref{prop-X_m}, we have\\
{\bf A.i)} $\displaystyle{\frac{d}{dL}f\left(X_m(\theta,L),\theta,L\right)}<0$,\\
{\bf A.ii)} $\displaystyle{\frac{d}{d\theta}f\left(X_m(\theta,L),\theta,L\right)}<0$.\\
{\bf B) Properties of $\mathbf{\textit{f}(\widetilde{\theta})}$}\\
{\bf B.i)} Let $(\theta, L)\in (0, +\infty)^2$ such that $\widetilde{\theta}=\widetilde{\theta}(\theta,L)>0$.  Then we have $\displaystyle{\frac{d}{d\theta}f\left(\widetilde{\theta}(\theta,L),\theta,L\right)}<0$.\\
{\bf B.ii)} For $\theta>\theta^*$ and $L>L_{02}(\theta)$, we have $\widetilde{\theta}>0$ and $\displaystyle{\frac{d}{dL}f\left(\widetilde{\theta}(\theta,L),\theta,L\right)}<0$.
\end{lemma}

\noindent {\bf Proof of Lemma \ref{lem-f-theta-L}}\\
{\bf A)} Let $\theta>0$ and $L> L_d(\theta)$. Using Proposition \ref{prop-X_m} ii), $f$ admits a non zero minimizer
$X_m\in
(2\theta/5,\theta)$.\\
{\bf Proof of A.i)} We have
$$\frac{d}{dL}f\left(X_m(\theta,L),\theta,L\right)=\partial_ Lf\left(X_m,\theta,L\right) +\partial_Xf\left(X_m,\theta,L\right)\partial_LX_m=-X_m^2<0,$$
where we have used (\ref{energyf}) and (\ref{caract-X_m}) to get the last equality.\\
{\bf Proof of A.ii)} We have
$$\frac{d}{d\theta}f\left(X_m(\theta,L),\theta,L\right)=\partial_\theta f\left(X_m,\theta,L\right) +\partial_ X f\left(X_m,\theta,L\right)\partial_\theta X_m=-\frac{1}{2}(\theta- X_m)^{-3/2}<0,$$
where again we have used (\ref{energyf}) and (\ref{caract-X_m}) to get the last equality.\\
{\bf B)} We recall that $\widetilde{\theta}=\widetilde{\theta}(\theta,L)$ is given in (\ref{D}).\\
{\bf Proof of B.i)} Let $(\theta, L)\in (0, +\infty)^2$ such that $\widetilde{\theta}=\widetilde{\theta}(\theta,L)>0$. We have
$$\frac{d}{d\theta}f\left(\widetilde{\theta}(\theta,L),\theta,L\right)=\frac{d}{d\theta}\left[L\left(\frac{1}{\alpha}-\widetilde{\theta}^2\right)\right]=-2L\widetilde{\theta}<0.$$
{\bf Proof of B.ii)} 
It is easy to check that 
$\widetilde{\theta}(\theta,L_{02}(\theta))=\alpha^{-1/2}>0 \mbox{ for } \theta\geq \theta^*$. 
Since $\partial_L\widetilde{\theta}>0$, we deduce that
\begin{equation}\label{theta-tilda>0}
\widetilde{\theta}>\alpha^{-1/2}>0 \quad \mbox{ for } \theta> \theta^* \mbox{ and } L>L_{02}(\theta).
\end{equation}
Now because $\widetilde{\theta}>0$, a direct computation shows that for $\theta> \theta^*$ and $L>L_{02}(\theta)$ we have
$$\frac{d}{dL}f\left(\widetilde{\theta}(\theta,L),\theta,L\right)=\frac{d}{dL}\left[L\left(\frac{1}{\alpha}-\widetilde{\theta}^2\right)\right]=\frac{1}{\alpha}-\widetilde{\theta}^2-4\alpha^2\frac{\widetilde{\theta}}{L^2}.$$
Using (\ref{theta-tilda>0}) we get the result. 

\hfill $\Box$

\subsection{Classification of global minimizers of $E$}\label{sec-proof-prop}

\noindent In this subsection, we prove Theorem \ref{theo-class-blisters}.

\noindent {\bf Proof of Theorem \ref{theo-class-blisters}}\\
Using Theorem \ref{theo-description}, a minimizer
$\zeta=(\zeta_1,\zeta_2)$ of the energy $E$ is always defined as in
(\ref{zeta1-zeta2}) and (\ref{beta-T-A}). So we have to identify the value of $K\in\mathcal{D}$ solving Problem (\ref{aux-min-f}) in each case.

\noindent \textbf{Case A: $(\theta, L)\in D_0$}\\ 
{\bf Case A.i):} $\theta>0$ and $L\leq L_d(\theta)$\\ 
Using Proposition \ref{prop-X_m} i) we deduce that 
$$\underset{\mathcal{D}}{\operatorname{argmin}}f=\{0\} .$$

\noindent {\bf Case A.ii):} $\theta>0$ and $L_d(\theta)< L <L_{01}(\theta)$\\ 
Using Proposition \ref{prop-X_m} ii) we deduce that $f$ admits a non zero minimizer
$X_m\in
(2\theta/5,\theta)$. Using (\ref{K-K-bar}), a simple computation leads us to the following
\begin{equation}\label{f(Xm)=f(0)}
f(X_m)=0=f(0) \quad \mbox{ for } \theta>0 \mbox{ and } L=L_{01}(\theta).
\end{equation}
Using Lemma \ref{lem-f-theta-L} A.i), we deduce that $f(X_m)>0=f(0)$ for $\theta>0$ and $L_d(\theta)< L <L_{01}(\theta)$.
Then
$$\underset{\mathcal{D}}{\operatorname{argmin}}f=\{0\}.$$

\noindent {\bf Case A.iii):} $\theta>\theta^*$ and $L_{01}(\theta)\leq L <L_{02}(\theta)$\\ 
Using Lemma \ref{rem-position}, we have $L\geq L_{01}(\theta)>L_d(\theta)$. Now using Proposition \ref{prop-X_m} ii), we deduce that
 $f$ admits a non zero minimizer
$X_m\in
(2\theta/5,\theta)$. 
Using Lemma \ref{lem-Xm-theta-tilda} ii), we conclude that
\begin{equation}\label{Xm>theta-tilda-L=L01}
X_m>\widetilde{\theta} \quad \mbox{ for } \theta>\theta^* \mbox{ and } L=L_{01}(\theta).
\end{equation}
Using Lemma \ref{lem-Xm-theta-tilda} i), we deduce that
$X_m>\widetilde{\theta}$ for $\theta>\theta^*$ and $L_{01}(\theta)\leq L <L_{02}(\theta)$. Therefore $\underset{\mathcal{D}}{\operatorname{argmin}}f\in\left\{0, \widetilde{\theta}\right\}$. Now, we distinguish two cases:\\
If $\widetilde{\theta}\leq 0$, then $\mathcal{D}=\{0\}$ and $\underset{\mathcal{D}}{\operatorname{argmin}}f=\{0\} $.\\ 
If $\widetilde{\theta}> 0$, we proceed as follows:
A direct computation shows that
\begin{equation}\label{f(theta-tilda)=f(0)}
f(\widetilde{\theta})=0=f(0)\quad \mbox{ for } \theta>\theta^* \mbox{ and } L=L_{02}(\theta).
\end{equation}
According to Lemma \ref{lem-f-theta-L} B.i), we deduce that $f(\widetilde{\theta})>0=f(0)$ for $\theta>\theta^*$ and $L_{01}(\theta)\leq L <L_{02}(\theta)$. Therefore $\underset{\mathcal{D}}{\operatorname{argmin}}f=\{0\} $.\\
This shows that in case A.iii), we have 
$$\underset{\mathcal{D}}{\operatorname{argmin}}f=\{0\} .$$

\noindent \textbf{Case B: $(\theta, L)\in D_1$}\\ 
According to Lemma \ref{rem-position}, we have $L>L_{01}(\theta)>L_d(\theta)$ for $0<\theta\leq \theta^*$. Using again Lemma \ref{rem-position}, we also have $L>L_{12}(\theta)>L_{01}(\theta)>L_d(\theta)$ for $\theta> \theta^*$. Therefore, we have $L>L_d(\theta)$ for all $(\theta,L)\in D_1$. Then using Proposition \ref{prop-X_m} ii), we deduce that
 $f$ admits a non zero minimizer
$X_m\in
(2\theta/5,\theta)$. 
Using Lemma \ref{lem-Xm-theta-tilda} ii), we conclude that
\begin{equation}\label{Xm<theta-tilda-L=L01}
X_m<\widetilde{\theta} \quad \mbox{ for } 0<\theta<\theta^* \mbox{ and } L=L_{01}(\theta).
\end{equation}
Moreover using Lemma \ref{lem-Xm-theta-tilda} iii), we have
\begin{equation}\label{Xm=theta-tilda}
X_m=\widetilde{\theta} \quad \mbox{ for } \theta\geq\theta^* \mbox{ and } L=L_{12}(\theta).
\end{equation}
Using Lemma \ref{lem-Xm-theta-tilda} i), we deduce that $X_m<\widetilde{\theta}  \mbox{ for } (\theta,L)\in D_1$. Therefore $\underset{\mathcal{D}}{\operatorname{argmin}}f\in\left\{0, X_m\right\}$.
Using (\ref{f(Xm)=f(0)}) and Lemma \ref{lem-f-theta-L} A.ii), we get that $f(X_m)<0=f(0) \mbox{ for } (\theta,L)\in D_1$. Therefore
$$\underset{\mathcal{D}}{\operatorname{argmin}}f=\{X_m\} .$$

\noindent \textbf{Case C: $(\theta,L)\in D_2 $}\\ 
According to Lemma \ref{rem-position}, we have $L>L_{02}(\theta)>L_{01}(\theta)>L_d(\theta)$ for $\theta>\theta^*$. Then using Proposition \ref{prop-X_m} ii), we deduce that
 $f$ admits a non zero minimizer
$X_m\in
(2\theta/5,\theta)$. 

\noindent {\bf i)} For $\theta> \theta^*$ and $L^*< L <L_{12}(\theta)$, 
using (\ref{Xm=theta-tilda}) and Lemma \ref{lem-Xm-theta-tilda} i), we deduce that $X_m>\widetilde{\theta}$. 

\noindent {\bf ii)} For $\theta> \theta^*$ and $L_{02}(\theta)< L\leq L^*$,  using (\ref{Xm>theta-tilda-L=L01}) and Lemma \ref{lem-Xm-theta-tilda} i), we deduce that $X_m>\widetilde{\theta}$. \\
Therefore for $(\theta, L) \in D_2$, we have $\underset{\mathcal{D}}{\operatorname{argmin}}f\in\left\{0, \widetilde{\theta}\right\}$.\\
On the other hand, using Lemma \ref{lem-f-theta-L} B.ii) for $\theta>\theta^*$ and $L>L_{02}(\theta)$, we have $\widetilde{\theta}>0$.  Using (\ref{f(theta-tilda)=f(0)}) and Lemma \ref{lem-f-theta-L} B.ii), we deduce that $f(\widetilde{\theta})<0=f(0) $ for $(\theta, L) \in D_2$.
 Therefore
$$\underset{\mathcal{D}}{\operatorname{argmin}}f=\{\widetilde{\theta}\} .$$

\noindent \textbf{Case D: $(\theta, L)\in \Gamma_{01}$}\\  
It is easy to check that 
$$\underset{\mathcal{D}}{\operatorname{argmin}}f=\{0, \overline{X}\}  \quad \mbox{ with } \quad 0<\overline{X}=X_m<\widetilde{\theta}.$$

\noindent \textbf{Case E: $(\theta, L)\in \Gamma_{02}$}\\ 
It is easy to verify that
$$\underset{\mathcal{D}}{\operatorname{argmin}}f=\{0, \overline{X}\}  \quad \mbox{ with } \quad 0<\overline{X}=\widetilde{\theta}<X_m.$$

\noindent \textbf{Case F: $(\theta, L)\in \Gamma_{12}$}\\ 
Similarly, we can show that
$$ \underset{\mathcal{D}}{\operatorname{argmin}}f=\{\overline{X}\}  \quad \mbox{ with } \quad \overline{X}=X_m=
\widetilde{\theta}>0.$$

\noindent \textbf{Case G: $(\theta, L)=P=(\theta^*, L^*)$}\\
Finally, we can check that
$$ \underset{\mathcal{D}}{\operatorname{argmin}}f=\{0, \overline{X}\}  \quad \mbox{ with } \quad 0<\overline{X}=X_m=\widetilde{\theta}.$$

\noindent {\bf Conclusion:} 
So we have proved that
\begin{equation}\label{unique-K}
\underset{\mathcal{D}}{\operatorname{argmin}}f=\left\{\begin{array}{ll}
\{0\}  \quad  &\mbox{ if } \quad (\theta,L)\in D_0; \\
\\
\{X_m\}  \quad  &\mbox{ if } \quad (\theta,L)\in D_1;\\
\\
\{\widetilde{\theta}\}  \quad  &\mbox{ if } \quad (\theta,L)\in D_2\cup\Gamma_{12};
\end{array}\right.
\end{equation}
and
\begin{equation}\label{double-K}
\underset{\mathcal{D}}{\operatorname{argmin}}f\in\left\{\begin{array}{ll}
\{0,X_m\}  \quad  &\mbox{ if } \quad (\theta,L)\in\Gamma_{01};\\
\\
\{0,\widetilde{\theta}\}  \quad  &\mbox{ if } \quad (\theta,L)\in \Gamma_{02}\cup\{P\}.
\end{array}\right.
\end{equation} 
Now using Theorem \ref{theo-description}, a minimizer
$\zeta=(\zeta_1,\zeta_2)$ of the energy $E$ is defined as in
(\ref{zeta1-zeta2}) and (\ref{beta-T-A}). Moreover, using (\ref{T-K}) we get
$T<\overline{L}$ if $(\theta,L)\in D_1$, and $T=\overline{L}$ if
$(\theta,L)\in D_2\cup\Gamma_{12}$ which shows (\ref{T-domains}). Similarly, we get (\ref{T-boundary}) for $(\theta,L)\in
\Gamma_{01}$ or $(\theta,L)\in \Gamma_{02}\cup\{P\}$.

\hfill $\Box$

\subsection{Global minimizers of $E$ and blister's properties}
In this subsection, we prove Theorem \ref{theo-description} ii) and 
Proposition \ref{prop-property}.\\

\noindent {\bf Proof of Theorem \ref{theo-description} ii)}\\
\textbf{a) Case $(\theta, L)\in  D_0\cup D_1\cup D_2\cup \Gamma_{12}$}\\
Using
(\ref{unique-K}), there exits a unique minimizer $K\in
\mathcal{D}$ of the function $f$. Now using Theorems
\ref{theo-blisters} and \ref{theo-description} i), there exists a
global minimizer $\zeta\in Y$ of the energy $E$ and there exits
$T\in[0,\overline{L}]$ such that (up to addition of constants and
translation of $(\zeta_1,\zeta_2)$) this minimizer is given by
(\ref{zeta1-zeta2}) and (\ref{beta-T-A}).\\
\textbf{b) Case $(\theta, L)\in \Gamma_{01}\cup\Gamma_{02}\cup \{P\}$}\\
Using
(\ref{double-K}), there exit exactly two minimizers of the
function $f$ on $\mathcal{D}$. Similarly using (\ref{equ-E-f}), Theorems
\ref{theo-blisters} and \ref{theo-description} i), we see that the energy $E$
has exactly two global minimizers in $Y$: the trivial solution
$(\zeta_1,\zeta_2)=(0,0)$ and a blister $\zeta\in Y$ given by
(\ref{zeta1-zeta2})-(\ref{beta-T-A}). 

\hfill $\Box$

\noindent {\bf Proof of Proposition \ref{prop-property}}\\
{\bf Proof of i)}\\ 
Let $(\theta,L)\in D_1\cup D_2\cup\Gamma_{12}$. Using (\ref{unique-K}), we conclude that there exists a unique $K=K(\theta,L)\in\mathcal{D}$ (see (\ref{unique-K})) such that
$$\min_{X\in\mathcal{D}}f(X)=f(K),$$
with $f$ and $\mathcal{D}$, respectively introduced in (\ref{energyf}) and
(\ref{D}). According to (\ref{beta-T-A}), the length of the support of
$\zeta_2$ and its amplitude $A$ can be written (using (\ref{a-b}) to express $(\overline{\theta},\overline{L})$ in terms of $(\theta,L)$) as
\begin{equation}\label{T-proof}
T=2\pi \sqrt{\frac{2\alpha}{3(\overline{\theta}-\alpha K)}}=2\pi\sqrt{\frac{2}{3}}(\theta-K)^{-1/2},
\end{equation}
and
\begin{equation}\label{A-proof}
A=\sqrt{\frac{K\overline{L}}{\pi\beta}}=\left(\frac{2}{3}\right)^{1/4}(\pi\alpha)^{-1/2}\left(KLT\right)^{1/2}.
\end{equation}
Using (\ref{unique-K}), we have
\begin{equation*}
K=K(\theta,L)=\left\{\begin{array}{ll}
X_m(\theta,L) \quad  &\mbox{ if } \quad (\theta,L)\in D_1;\\
\\
\widetilde{\theta}(\theta,L)  \quad  &\mbox{ if } \quad (\theta,L)\in D_2\cup\Gamma_{12}.
\end{array}\right.
\end{equation*}
The function $\widetilde{\theta}$ is smooth. Moreover, $X_m$ is smooth on $D_1$ according to Proposition \ref{prop-X_m} iii). Therefore
 $K=K(\theta,L)$ is a smooth function of $(\theta,L)$ on each domain $D_1$ and $D_2$. \\
Straightforward calculations show that for $(\theta,L)\in D_1\cup D_2$
\begin{equation}\label{derive-A-T}
\left\{\begin{array}{ll} \displaystyle{\partial _{\theta}T}&=\displaystyle{\pi \sqrt{\frac{2}{3}}(\theta-K)^{-3/2}(\partial_{\theta}K-1)},\\
\\
\displaystyle{\partial _{L}T}&=\displaystyle{\pi\sqrt{\frac{2}{3}} (\theta-K)^{-3/2}\partial_{L}K},\\
\\
\displaystyle{\partial _{\theta}A}&=\displaystyle{\left(\frac{1}{24}\right)^{1/4}\left(\frac{L}{\alpha\pi K T}\right)^{1/2}(T\partial_{\theta}K+K\partial_{\theta}T)},\\
\\
\displaystyle{\partial _{L}A}&=\displaystyle{\left(\frac{1}{24}\right)^{1/4}\left(\frac{1}{\alpha\pi
K T L}\right)^{1/2}(KT+TL\partial_{L}K+KL\partial_{L}T)}.
\end{array}\right.
\end{equation}
\textbf{Case 1: $ (\theta, L)\in D_1$}\\ 
Using (\ref{deriv-K-theta}), (\ref{deriv-K-L}) and (\ref{derive-A-T}), we conclude that
$\partial_{\theta}T, \partial_{L}T, \partial_{\theta}A, \partial_{L}A\geq 0$.\\
\textbf{Case 2: $(\theta, L)\in D_2\cup\Gamma_{12}$}\\ 
Using (\ref{unique-K}), we have
$K=\widetilde{\theta}(L)=\theta-\alpha^2/L^2$ and $T=\overline{L}$.
So we have
$$\partial_{\theta}K=1 \quad \mbox{ and } \quad \partial_{L}K=\frac{2\alpha^2}{L^3}>0.$$
Using (\ref{derive-A-T}), we conclude that
$\partial_{\theta}T, \partial_{L}T, \partial_{\theta}A, \partial_{L}A\geq 0$.\\
{\bf Proof of ii)}\\
{\bf Step 1: Proof of (\ref{T-star})}\\
Our goal is to compute the derivative of $T$ with respect to $\theta$ along the curve $\Gamma_{01}$. Using (\ref{derive-A-T}), (\ref{deriv-K-theta}) and (\ref{deriv-K-L}), we get with obvious notation for $(\theta,L)\in \Gamma_{01}$ (using the fact that $L=L_{01}(\theta)$ given in (\ref{b+}))
\begin{eqnarray*}
\frac{d}{d\theta}T(\theta, L_{01}(\theta))&=&\pi\sqrt{\frac{2}{3}}(\theta-K)^{-3/2}[\left(\partial_{\theta}K-1\right) +(\partial_{L}K) L'_{01}(\theta)]\\
&=&\pi\sqrt{\frac{2}{3}}(\theta-K)^{-3/2}\left[\frac{5^{5/2}\theta^{-5/2}(2-5\theta^{-1}K)}{16\left(3/4(\theta-K)^{-5/2}-2L\right)}\right].
\end{eqnarray*}
Using (\ref{caract-X_m}) and (\ref{f''(K)}), we conclude that $T=T(\theta, L_{01}(\theta))$ is decreasing in $\theta$ along the curve $\Gamma_{01}$.
Then we deduce that $\underset{(\theta, L)\in \Gamma_{01}}{\inf}T=T(P)=T(\theta^*,L_{01}(\theta^*))$. Using (\ref{T-boundary}), we have $T(P)=\overline{L}^*$, where $\overline{L}^*$ is the value of $\overline{L}$ at the point $P$. Therefore
\begin{equation}\label{T*}
\inf_{(\theta, L)\in \Gamma_{01}}T=\overline{L}^*=2\pi\sqrt{\frac{2}{3}}\frac{L_{01}(\theta^{*})}{\alpha}=4\pi \sqrt{\frac{2}{3}}\alpha^{1/4}=:T^*.
\end{equation}
Using the monotonicity of $T$ in $\theta$ and $L$ on $D_1$, we get (\ref{T-star}).


\noindent{\bf Step 2: Proof of (\ref{A-star})}\\
Let $\overline{K}:=K/\theta$. Similarly  using (\ref{A-proof}) and (\ref{T-proof}), we explicit $A$ in term of $\overline{K}$ for $(\theta,L)\in \Gamma_{01}$ (in particular $L=L_{01}(\theta)$). 
A straightforward computation gives
$$A(\theta,L_{01}(\theta))=\left(\frac{1}{12}\right)^{1/2}5^{5/4}\left(\overline{K}(1-\overline{K})^{-1/2}\right)^{1/2}\alpha^{-1/2}\theta^{-1}.$$
For $(\theta, L)\in \Gamma_{01}$, with $K=X_m$ we have by (\ref{K-K-bar}) that $\overline{K}=4/5$.
Then $A=A(\theta,L_{01}(\theta))$ is decreasing in $\theta$ along the curve $\Gamma_{01}$. So
$$
\inf_{(\theta, L)\in \Gamma_{01}}A=A(\theta^{*},L_{01}(\theta^*))=2\left(\frac{5}{9}\right)^{1/4}\left(\overline{K}(1-\overline{K})^{-1/2}\right)^{1/2}=4/\sqrt{3}=:A^{*}.
$$
Finally using the monotonicity of $A$ in $\theta$ and $L$ on $D_1$, we get (\ref{A-star}).

\hfill $\Box$

\section*{Acknowledgment}

The authors thank A. El Doussouki for helpful discussions on the problem.
This work has been financially supported by the project CEDRE 11EF45 L20 (2012-2013).



\begin{thebibliography}{99}

\bibitem{A2}
{\sc B. Audoly,}
{\it Stability of Straight Delamination Blisters,}
J. of Physical Review Letters, 83 (20) (1999), 4124-4127.

\bibitem{B-B-M-M}
{\sc A. A. L. Baldelli, B. Bourdin,  J. J. Marigo, C. Maurini,}
{\it Delamination and fracture of thin films: a variational approach,}
Direct and variational methods for nonsmooth problems in mechanics
Amboise, France, June 24-26, (2013).


\bibitem{B-K}
{\sc J. Bedrossian, R. V. Kohn,}
{\it Blister patterns and energy minimization in compressed thin films on compliant substrates,}
Preprint (2013).

\bibitem{C}
{\sc G. Chmaycem,}
PhD Thesis, Ecole Nationale des Ponts et Chauss\'ees, (in preparation). 

\bibitem{D}
{\sc A. El Doussouki,}
PhD Thesis, Universit\'e de Picardie Jules Vernes, (2012). 

\bibitem{F-M}
{\sc G. Francfort, J.J. Marigo,}
{\it Griffith Theory Of Brittle Fracture Revisited: Merits And Drawbacks,}
J. of Latin American Journal of Solids and Structures, Vol. 2 (2005),  57-64.

\bibitem{G}
{\sc A. A. Griffith,}
{\it The Phenomena of Rupture and Flow in Solids,}
J. of Philosophical Transactions of the Royal Society of London. Series A, Containing Papers of a Mathematical or Physical Character, Vol. 221 (1920), 163-198. 



\bibitem{G-O}
{\sc G. Gioia, M. Ortiz,}
{\it Delamination of Compressed Thin Films,}
J. of Advances in Applied Mechanics, 33 (1997), 119-192.

\bibitem{L}
{\sc C. J. Larsen,}
{\it Models of dynamic fracture based on Griffith's criterion,}
IUTAM Symposium on Variational Concepts with Applications to the Mechanics of Materials: Proceedings of the IUTAM Symposium on Variational Concepts with Applications to the Mechanics of Materials, Bochum, Germany, September 22-26, (2008), Vol. 21. Springer, (2010).


\bibitem{M}
{\sc R. Monneau,}
{\it Some remarks on the asymptotic invertibility of the linearized operator of nonlinear elasticity in the context of the displacement approach,}
J. of ZAMM  Z. Angew. Math. Mech., (2006), 1-10. 


\bibitem{V-B-B-R-R}
{\sc D. Vella, J. Bico, A. Boudaoud, B. Roman, P. M. Reis,}
{\it The macroscopic delamination of thin films from elastic substrates,}
J. of Proceedings of the National Academy of Sciences 106.27 (2009), 10901-10906.

\bibitem{V-W}
{\sc A. A. Volinsky, P. Waters,}
{\it Delaminated Film Buckling Microchannels,}
J. of Mechanical Self-Assembly: Sciences and Applications, (2013), 153-170.







\end{thebibliography}
\end{document}